\documentclass[prl,a4paper,amsfonts,amssymb,onecolumn,showpacs,showkeys,nofootinbib,floats]{revtex4}

\usepackage{amsfonts}
\usepackage{amsmath}
\usepackage{amssymb}
\usepackage{graphicx}

\begin{document}

\title[Traded volume dynamics in finance]{On the distribution of high-frequency stock market traded volume: 
a dynamical scenario}

\author{S\'{\i}lvio M. Duarte Queir\'{o}s}

\affiliation{Centro Brasileiro de Pesquisas F\'{\i}sicas, 150, 22290-180, Rio de Janeiro - RJ, Brazil\\ 
{\it Electronic mail address:}{ \tt sdqueiro@cbpf.br}\\
}

\date{\today}

\begin{abstract}
This manuscript reports a stochastic dynamical scenario whose associated
stationary probability density function is exactly a previously proposed one 
to adjust high-frequency traded volume distributions. This dynamical conjecture, physically 
connected to superstatiscs, which is intimately related with
the current nonextensive statistical mechanics framework, is based on the idea of 
local fluctuations in the mean traded volume associated to financial markets agents herding behaviour. 
The corroboration of this mesoscopic model is done by modelising NASDAQ 1 and 2 minute stock market
traded volume.
\end{abstract}



\maketitle

%\begin{multicols}{1}

In the last years we have assisted to a rapidly growing interesting of
physicists in the analysis of financial systems and the description of their
statistical and dynamical properties~\cite{bouchaud-potters,mantegna-stanley-book,farmer,voit,gm-ct}. 
Actually, some of the so-called \textit{stylised facts} such as non-Gaussianity in relative price change 
(the \textit{return}) probability density functions
(PDFs)~\cite{mantegna-stanley-nature,plerou-amaral-stanley,inpe,ausloos-ivanova,osorio-borland-tsallis,smdq-lisboa}, 
multi-fractal behaviour of return time series~\cite{ghashghaie,bouchaud-potters-meyer,arneodo-sornette}, 
asymptotic power-law like auto-correlation funtion in the
standard deviation (the \textit{volatility}) of return~\cite{ding-granger-engle,potters-cont-bouchaud}, 
among many others, were (also) verified in a physical context. Besides their own relevance, 
these empirical observations were important to the establishment of either new theoretical models, which mimic
reality much better than previous ones~\cite{potters-cont-bouchaud,gabaix-stanley,farmer-daniels,farmer-lillo}, 
or connections between financial systems and already existing
statistical frameworks~\cite{osorio-borland-tsallis,borland-prl,smdq-quantf},
particularly, the current non-extensive statistical mechanics~\cite{gm-ct},
based on \textit{Tsallis entropy}~\cite{ct}, which, {\it e.g.}, allowed the
derivation of a new and closed model for option-pricing~\cite{borland-quantf1,borland-bouchaud,borland-quantf2}.

\begin{figure}[tbp]
\begin{center}
\includegraphics[width=0.4\columnwidth,angle=0]{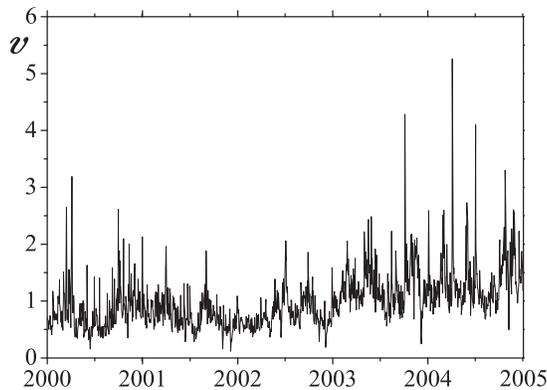}
\end{center}
\caption{Evolution of Microsoft Corporation daily normalised traded volume in NASDAQ exchange from the 
$21^{st}$ January $2001$ to the $21^{st}$ January $2005$. Each tick represents 6 months.}
\label{fig-1}
\end{figure}

Additionally to macroscopic observables as the return and the volatility,
mentioned above, another significant quantity in financial markets is the
\textit{traded volume}, $V$, which corresponds to the number of shares traded
(in a determined period of time, $T$). This is, in fact, along with
volatility, an important parameter in trading strategies assumed by many
brokers~\cite{willmott,ausloos-ivanova2}. In fig.~\ref{fig-1} is presented, as mere
illustration, the daily normalised traded volume of Microsoft Corp. in
NASDAQ exchange from $2000$ to the beginning of $2005$.

In a previous work, P. Gopikrishnan \textit{et al.}~\cite{gopi} reported the
evidence, for large volumes, of a power-law behaviour in the stationary
PDF. The analysis of traded volume stationary PDF was later extended to all 
range of values by R. Osorio, L. Borland and C. Tsallis (OBT)~\cite{osorio-borland-tsallis}, 
who proposed a fitting expression that presents an amazing precision. 
The respective formula is,
\begin{equation}
p\left(  v\right)  =\frac{1}{Z}\left(  \frac{v}{^{\theta}}\right)  ^{\alpha
}\exp_{q}\left(  -\frac{v}{\theta}\right)  , \label{prob-obt}
\end{equation}
where $v$ represents the traded volume expressed in its mean value unit ($\left\langle 
V\right\rangle $), {\it i.e.}, $v_{t} = \frac{V_{t}}{\left\langle V\right\rangle}$
$Z=\int\nolimits_{0}^{\infty}\left(  \frac{v}{^{\theta}}\right)^{\alpha}\exp_{q}\left(  -\frac{v}{\theta}\right)  dv$, 
$\alpha $ and $\theta $ are positive parameters and
\begin{equation}
\exp_{q}\left(  x\right)  \equiv\left[  1+\left(  1-q\right)  \,x\right]
^{\frac{1}{1-q}},\qquad(\exp_{1}\left(  x\right)  =e^{x})  , \label{q-exp}
\end{equation}
is the $q$\textit{-exponential function}, which emerged in the context of the 
\textit{non-extensive statistical mechanics}~\cite{gm-ct}.

In the sequel of this article, through a set of liable arguments, we will introduce a stochastic dynamical
mechanism, which has eq.~(\ref{prob-obt}) as stationary solution for its probability density
function and then compare both analytical and simulation results using values obtained from NASDAQ stock market.

\medskip

Consider the following stochastic differential equation,
\begin{equation}
dv=-\gamma\,\left(  v-\frac{\alpha+1}{\beta}\right)  \,dt+\sqrt{2\,v\,\frac
{\gamma}{\beta}}\,dW_{t},\label{equacao1}
\end{equation}
($W_{t}$ represents a regular Wiener process following a normal
distribution with null mean and unitary variance) as the traded volume updating law~\footnote{Eq.~(\ref{equacao1}) 
is in fact defined in mathematics as {\it Feller process}~\cite{feller}.According to its definition the traded 
volume $V$ is an integer quantity. However, the normalised traded volume is a 
rational quantity and so perfectly compatible with a continuous approach.}. 
As can be seen, eq.~(\ref{equacao1}) is composed by a deterministic term, similar to a restoring force and a 
stochastic term, which is a typical example of multiplicative noise very usual in problems presenting clustering 
profiles, like the one treated in this manuscript (see fig.~\ref{fig-1}), intermittence~\cite{takayasu-sato-takayasu} 
or turbulence~\cite{frisch,beck-lewis-swinney}.
The first term can be interpreted as a restoring market mechanism (with constant $\gamma $), 
which aims to keep the traded volume of the stock at some typical value $\Theta = \frac{\alpha+1}{\beta}$.~\footnote{
Following $v$ definition, $\Theta $ should equals $1$. Nonetheless, we will keep it in the presented form for reasons 
that will become clear later on.} 
The second one reflects stochastic memory, and mainly, the upshot of large traded volumes. In effect, large traded volumes 
will induce large values in the stochastic term, but, due to $W_{t}$ random character(sign), it 
could lead to an increase ($W_{t} > 0$) or a decrease ($W_{t} < 0$) of $v$, {\it i.e.}, to a stirred or serene 
state of the stock.

For eq.~(\ref{equacao1}) we can write the corresponding Fokker-Planck equation~\cite{risken},
\begin{equation}
\frac{\partial p(v,t)}{\partial t}=\frac{\partial}{\partial x}\left[
\gamma\,\left(  v-\frac{\alpha+1}{\beta}\right)  p(v,t)\right]  +\frac
{\partial^{2}}{\partial x^{2}}\left[  v\,\frac{\gamma}{\beta}\,p(v,t)\right]
,\label{fokker-planck1}
\end{equation}
and determining its stationary solution we obtain,
\begin{equation}
p\left(  v\right)  =\frac{\beta}{\Gamma\left[  1+\alpha\right]  }\left(
\beta\,v\right)  ^{\alpha}\,\exp\left(  -\beta\,v\right)  \qquad
(\alpha>-1,\ \beta>0),\label{p1v}
\end{equation}
which is just the Gamma distribution, with mean value,
$\left\langle v\right\rangle =\frac{1+a}{\beta}$, standard deviation,
$\left\langle v-\left\langle v\right\rangle \right\rangle ^{2}=\frac
{1+a}{\beta^{2}}$ and most probable value, $v^{\ast}=\frac{\alpha}{\beta}$
($\left.  \frac{dP\left(  v\right)  }{dv}\right\vert _{v=v^{\ast}}=0$).

Assume now that, instead of constant in time, $\beta$ is now randomly varying
on a time scale much larger than the time scale of order $\gamma^{-1}$
necessary by eq.~(\ref{fokker-planck1}) to reach stationarity, hypothesis that
is equivalent to consider local fluctuations in the mean value of
$v$~\cite{smdq-wp}, akin to the proposal of moving average in the volatility~\cite{perello}. 
These fluctuations can be microscopically associated to 
changes in the {\it volume of activity} defined as the number of traders who 
performed transactions of the stock at that time~\cite{lux}. 
Furthermore, let us assume that $\beta$ follows a Gamma (stationary) PDF,
\begin{equation}
P\left(  \beta\right)  =\frac{1}{\lambda\,\Gamma\left[  \delta\right]
}\left(  \frac{\beta}{\lambda}\right)  ^{\delta-1}\exp\left(  -\frac{\beta
}{\lambda}\right)  \qquad(\delta>0,\ \lambda>0).\label{p1b}
\end{equation}
This assumption transforms eq.~(\ref{p1v}) meaning from the absolute
probability of $v$ into the conditional probability of $v$ given $\beta$,
\begin{equation}
p\left(  v\right)  \rightarrow p\left(  v|\beta\right)  =\frac{\beta}%
{\Gamma\left[  1+\alpha\right]  }\left(  \beta\,v\right)  ^{\alpha}%
\,\exp\left(  -\beta\,v\right)  .\label{pvb}
\end{equation}
It is noteworthy that the problem of macroscopic non-equilibirium systems with
complex dynamics in stationary states with fluctuations of intensive
\ quantities on long time scales and its connection with non-extensive
statistical mechanics, based on entropy~\cite{ct},
\[
S_{q}=\frac{1-\sum_{i}\left[  p_{i}\right]  ^{q}}{q-1},
\]
($S_{1}=-\sum_{i}p_{i}\,\ln p_{i}\equiv S_{BG}$ where $BG$ stands for
Boltzmann-Gibbs) was firstly presented by G. Wilk and Z.
W\l odarczyk~\cite{wilk}, later applied by C. Beck to the case of turbulent
fluids~\cite{beck-prl} and finally extended and defined by C. Beck and E.G.D.
Cohen as \textit{superstatistics}~\cite{beck-cohen,ct-amcs}.

For this case, we have that the \textit{joint} probability of obtaining a certain
value of $v$ and certain value of $\beta$, is $P\left(  v,\beta\right)
=p\left(  v|\beta\right)  P\left(  \beta\right)  $ and the \textit{marginal}
probability of having some value $v$ is given by
\begin{equation}
P\left(  v\right)  =\int_{0}^{\infty}P\left(  v,\beta\right)  \,\,d\beta
=\int_{0}^{\infty}p\left(  v|\beta\right)  \,P\left(  \beta\right)
\,d\beta.\label{pv-int}
\end{equation}
Using eq.~(\ref{p1b}) and eq.~(\ref{pvb}) in eq.~(\ref{pv-int}) and performing
the integration we get,
\begin{equation}
P\left(  v\right)  =\frac{\lambda\,\Gamma\left[  1+\alpha+\delta\right]
}{\Gamma\left[  1+\alpha\right]  \,\,\Gamma\left[  \delta\right]  }\left(
\lambda\,v\right)  ^{\alpha}\,\left(  1+\lambda\,v\right)  ^{-1-\alpha-\delta
}\qquad(\alpha+\delta>-1).\label{pv2}
\end{equation}
Doing the change in parameters
\begin{equation}
\lambda=\frac{q-1}{\theta},\qquad\delta=\frac{1}{q-1}-\alpha
-1,\label{lambda-delta}%
\end{equation}
one can write eq.~(\ref{pv2}) as
\[
P\left(  v\right)  =\frac{\left(  q-1\right)  ^{\alpha+1}\,\Gamma\left[
\frac{1}{q-1}\right]  }{\theta\,\Gamma\left[  \frac{1}{q-1}-\alpha-1\right]
\,\,\Gamma\left[  \alpha+1\right]  }\left(  \frac{v}{\theta}\right)  ^{\alpha
}\,\left[  1+\left(  q-1\right)  \frac{v}{\theta}\right]  ^{1/\left(
1-q\right)  },
\]
or, using the $q$-exponential function definition, eq.~(\ref{q-exp}),
\begin{equation}
P\left(  v\right)  \equiv\frac{\left(  q-1\right)  ^{\alpha+1}\,\Gamma\left[
\frac{1}{q-1}\right]  }{\theta\,\Gamma\left[  \frac{1}{q-1}-\alpha-1\right]
\,\,\Gamma\left[  \alpha+1\right]  }\left(  \frac{v}{\theta}\right)  ^{\alpha
}\exp_{q}\left(  -\frac{v}{\theta}\right)  ,\label{pvfinal}
\end{equation}
which is \textit{exactly} the expression introduced by OBT. Due to its structure, PDF (\ref{pvfinal}) might 
be called $q$\textit{-generalised} \textit{Gamma probability density function}. 
The traditional Gamma PDF, eq.~(\ref{p1v}), can be recovered 
from eq.~(\ref{pvfinal}) by considering $q=1$
limit, which is equivalent to assume the inexistence of fluctuations in
$\beta$ and so $P\left(  \beta\right)  $ equals the Dirac delta function
centred in $\theta^{-1}$. For small values of $v$, $P\left(  v\right)  $ goes
to zero as a power law, $P\left(  v\right)  \sim v^{\alpha}$, and for large
values of $v$, it behaves asymptotically as another power law, $P\left(
v\right)  \sim v^{\alpha/(1-q)}$. As a matter of fact, this is an appealing form,
since it congregates, in a {\it single expression}, the two power laws used in
financial practice.

\medskip

\begin{figure}[tbp]
\begin{center}
\includegraphics[width=0.85\columnwidth,angle=0]{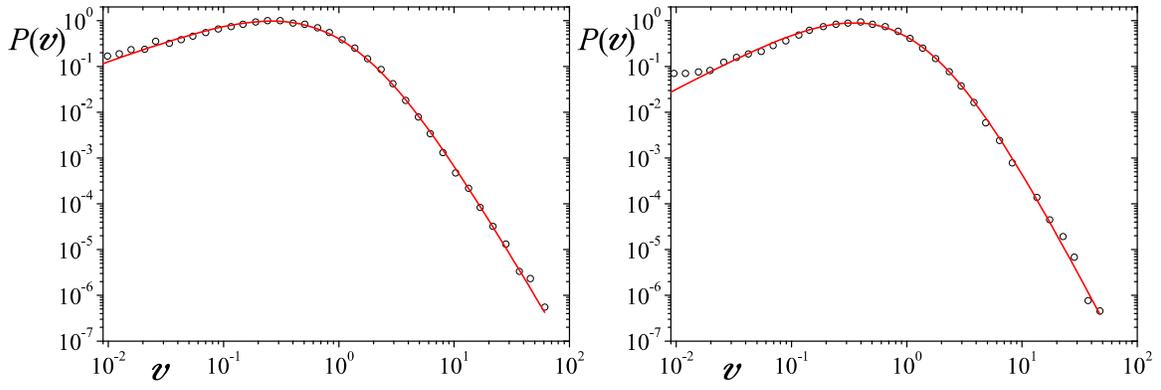}
\end{center}
\caption{1 minute (Left plot) and 2 minute (Right plot) traded volume probability density function $P(v)$, 
for the ten high-volume stocks in NASDAQ exchange during $2001$ {\it vs.} $v$. The symbols represent real data
and lines $p(v)$ from eq.(\ref{prob-obt}), with $q$, $\alpha $, $\theta $ values shown in table~\ref{tab-1}
(After OBT in ref.[10]). 
The $\chi^{2}$(per point) error function values are $1.8\times 10^{-3}$ and $6.5\times 10^{-4}$, respectively.
For very low volumes is visible a small discrepancy, due to lack of small transactions (statistical effect).}
\label{fig-2}
\end{figure}
In order to verify the correctness of the result, we performed a set of
numerical realisations, using an algorithm based on eq.~(\ref{equacao1}) and on
a random generator for $\beta$ whose associated probability is eq.~(\ref{p1b}).
The values of parameters $\alpha$, $\lambda$ and $\delta$, presented in
table~\ref{tab-1}, were obtained after fitting the $1$ minute and $2$ minute traded
volume probability density functions of the ten high-volume stocks in the
NASDAQ\ exchange during $2001$, fig.~\ref{fig-2}, with eq.~(\ref{prob-obt})
(equivalent to eq.~(\ref{pvfinal})). The values of $q$ and $\theta$ are
obtained using the relations presented in eq.~(\ref{lambda-delta}).
\begin{table}
\caption{Fitting parameters obtained in fig.\ref{fig-2} and dynamical parameters 
obtained through eq.\ref{lambda-delta}.}
\label{tab-1}
\begin{center}
\begin{tabular}{|c|c|c|c|c|c|}
\hline
& $q$ & $\alpha$ & $\theta$ & $\lambda$ & $\delta$\\
\hline\hline
$1\ \min$ & $1.19$ & $0.93$ & $0.23$ & $0.826$ & $3.33$\\ \hline
$2\ \min$ & $1.16$ & $1.36$ & $0.2$ & $0.8$ & $3.89$\\ \hline
\end{tabular}
\end{center}
\end{table}

\begin{figure}[tbp]
\begin{center}
\includegraphics[width=0.63\columnwidth,angle=0]{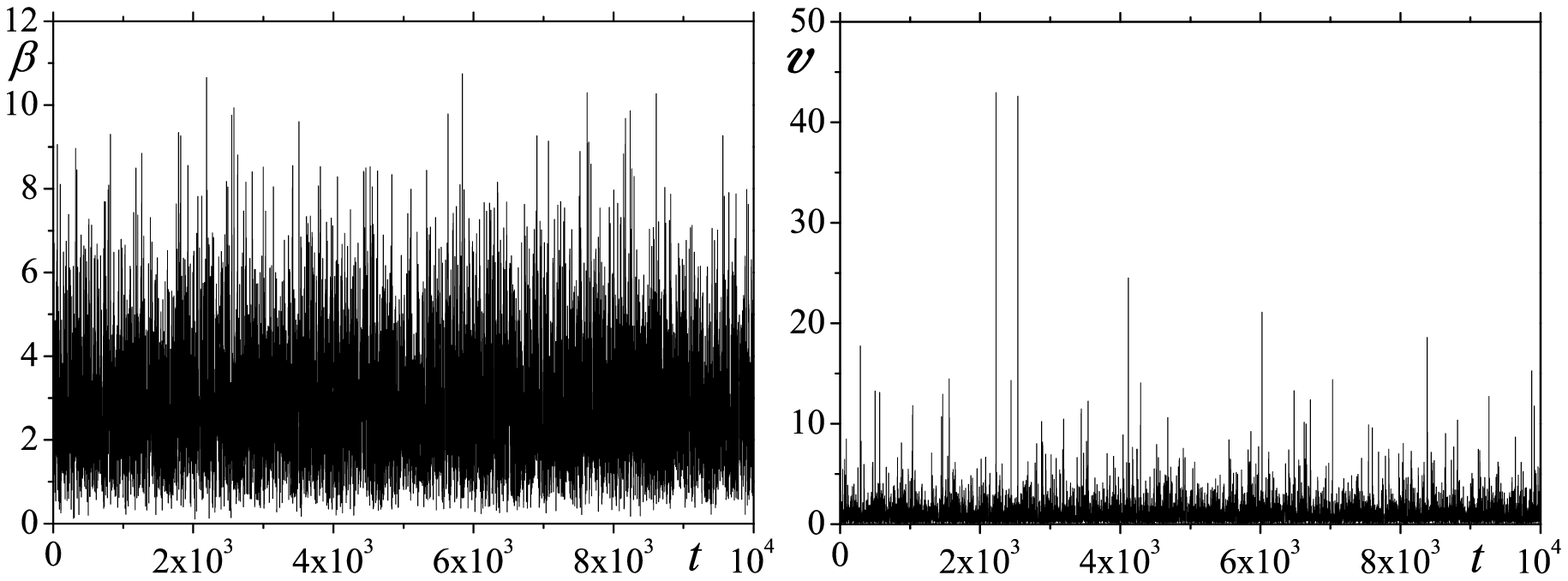}
\includegraphics[width=0.33\columnwidth,angle=0]{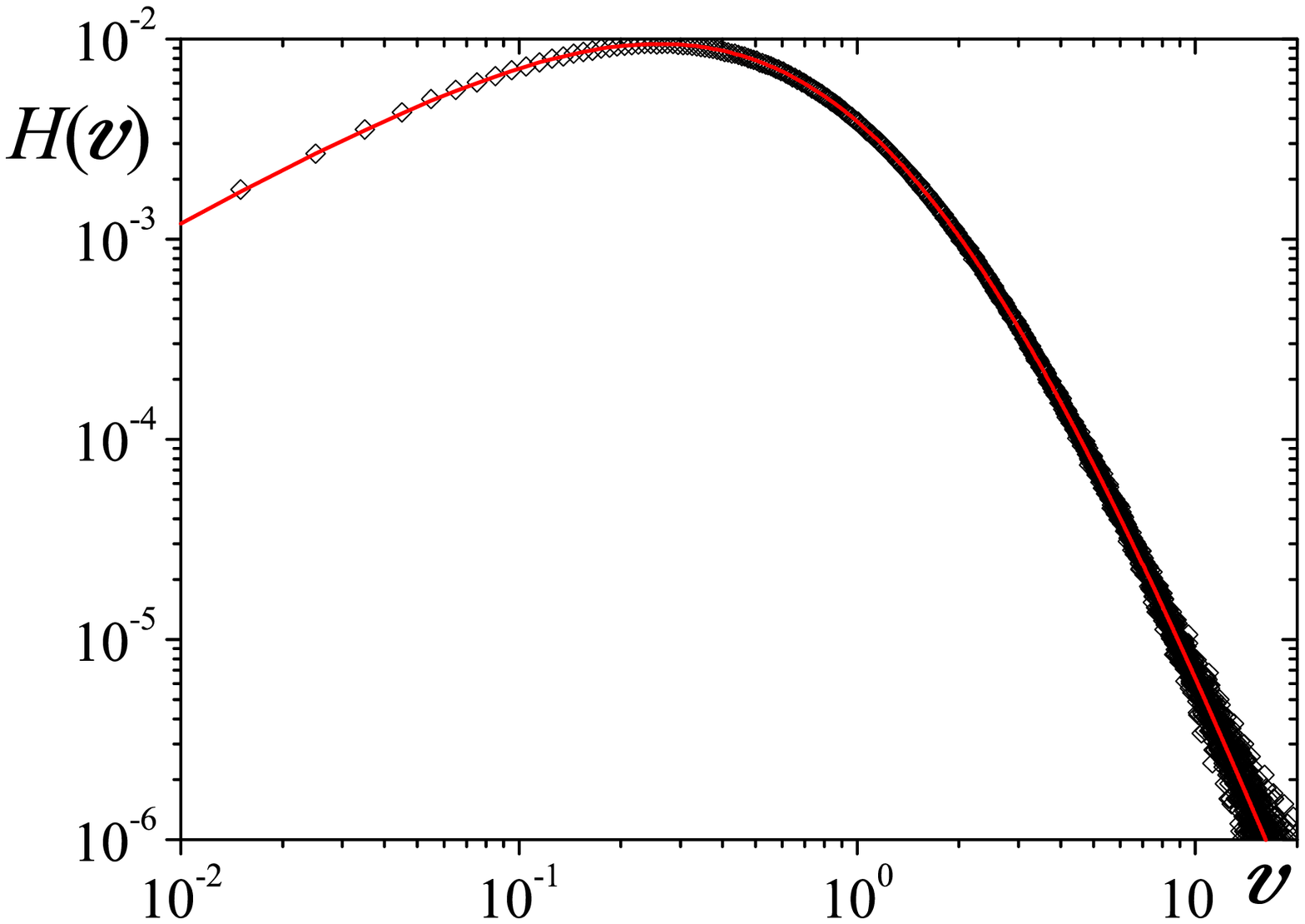}
\end{center}
\caption{Numerical realisation for $1$ minute traded volume of the ten high-volume stocks in NASDAQ exchange 
during $2001$ using the values of table~\ref{tab-1}. Left Plot: temporal evolution of $\beta$ parameter; 
Centre Plot: temporal evolution of traded volume, $v$. Right Plot: Relative frequency $H\left(  v\right)  $ 
\textit{vs.} $v$. The symbols were obtained numerically and the line from
$\int_{v-\Delta v}^{v+\Delta v}P\left(  v\right)  \,dv$ with $\Delta v=5\times10^{-3}$.
The ratio between $\beta$ updating time and stationarity
time scale of $v$ is $10^{2}$.}
\label{fig-3}
\end{figure}

\begin{figure}[tbp]
\begin{center}
\includegraphics[width=0.63\columnwidth,angle=0]{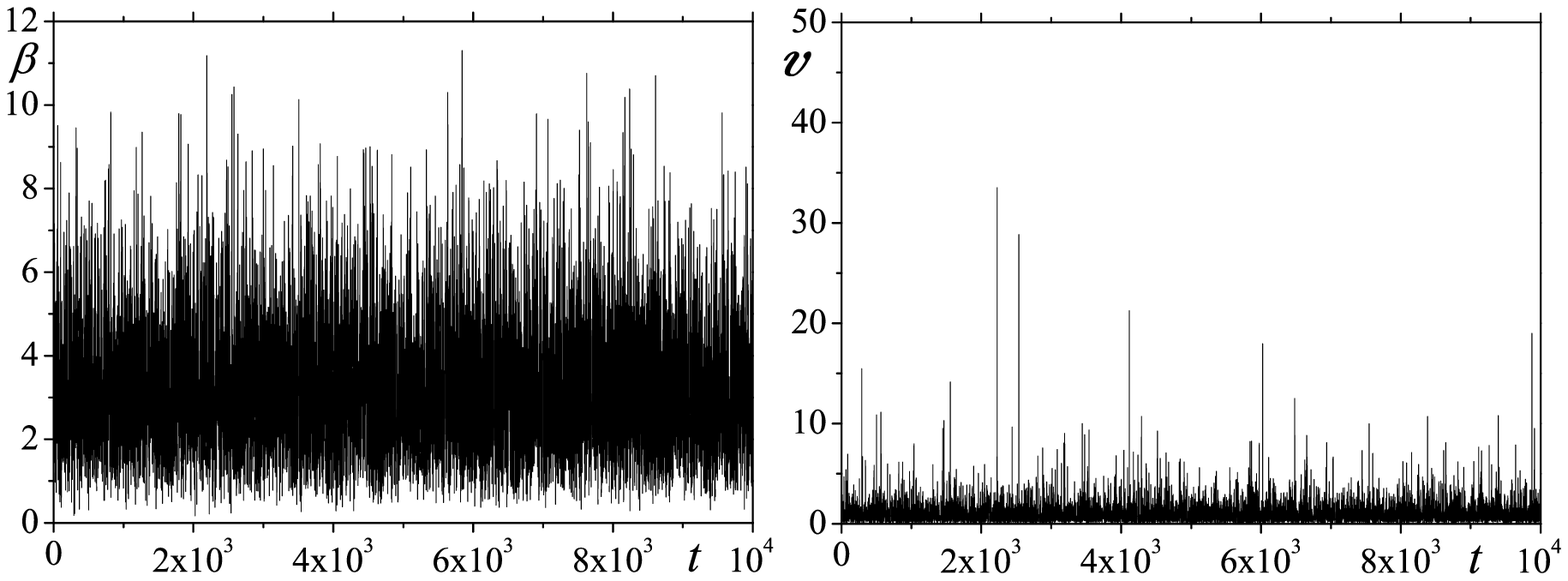}
\includegraphics[width=0.33\columnwidth,angle=0]{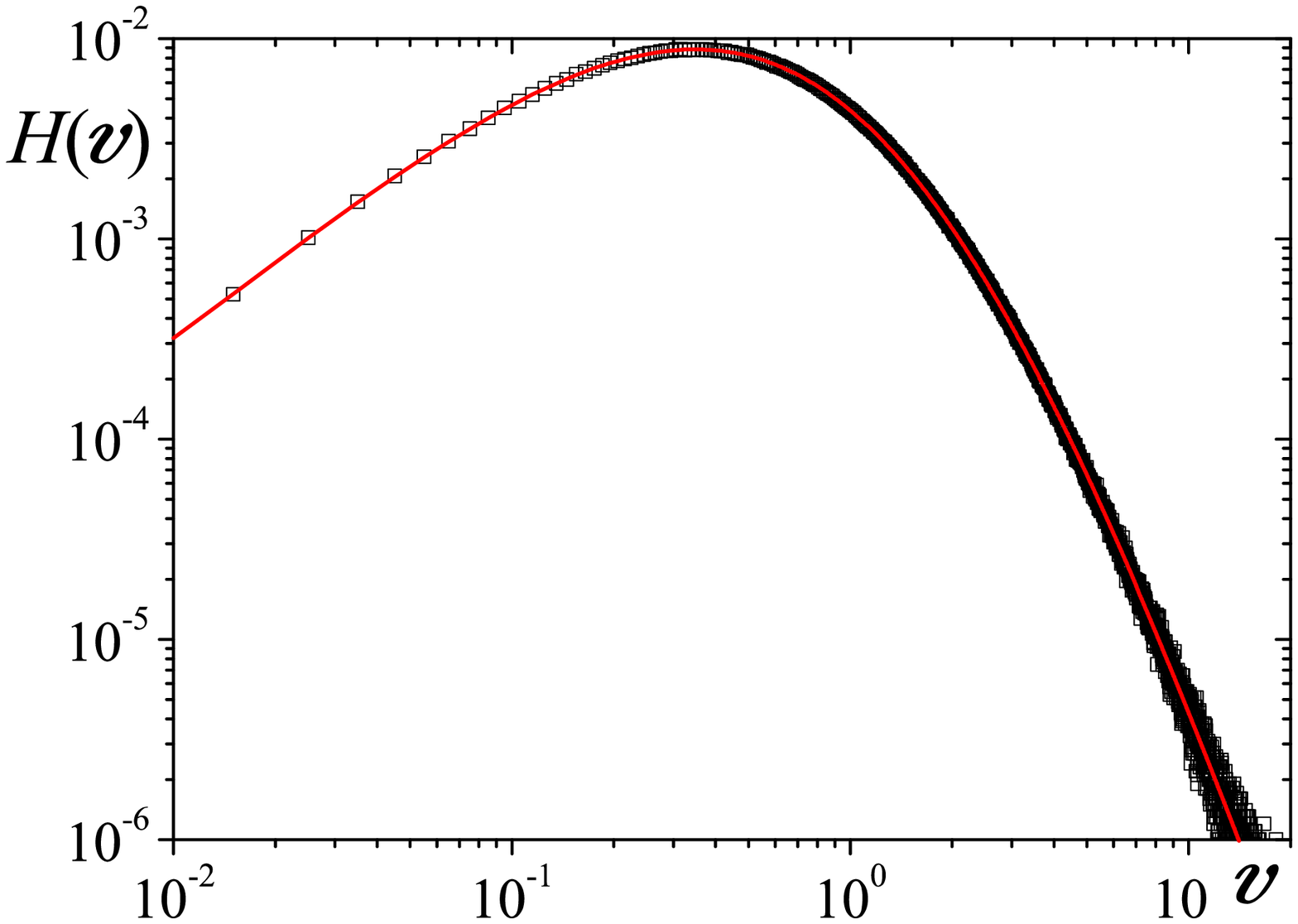}
\end{center}
\caption{Numerical realisation for $2$ minute traded volume of the ten high-volume stocks in NASDAQ exchange 
during $2001$ using the values of table~\ref{tab-1}. Left Plot: temporal evolution of $\beta$ parameter; 
Centre Plot: temporal evolution of traded volume, $v$. Right Plot: Relative frequency $H\left(  v\right)  $ 
\textit{vs.} $v$. The symbols were obtained numerically and the line from
$\int_{v-\Delta v}^{v+\Delta v}P\left(  v\right)  \,dv$ with $\Delta v=5\times10^{-3}$.
The ratio between $\beta$ updating time and stationarity
time scale of $v$ is $10^{2}$.}
\label{fig-4}
\end{figure}

The numerical realisations for $1$ minute and $2$ minute traded volume are
depicted in fig.~\ref{fig-3} and fig.~\ref{fig-4}.

In fig.~\ref{fig-1}, fig.~\ref{fig-3} and fig.~\ref{fig-4} (Centre plots) it is clear
the clustering behaviour, very similar to volatility time series, typical of a
multiplicative noise dynamics. In fig.~\ref{fig-3} and
fig.~\ref{fig-4} (Right plots), we can compare numerically obtained relative frequency, $H\left(v\right) $, 
with analitycal results corresponding to values presented in table~\ref{tab-1}. 
As it is straightforwardly visible, numerical and analytical results are in
perfect accordance. Besides the possibility of mimic traded volume time series, 
this dynamical conjecture can certainly be used in other problems with time varying 
positive quantities associated to a PDF (\ref{prob-obt}), such as granular material systems~\cite{kolb,ernesto}.

\medskip

To summarise, in this manuscript we presented a dynamical conjecture, based
on a multiplicative noise mechanism and on the fluctuating character of some
parameter of that system, which is able to provide the stationary probability
density function previously introduced to shape the high-frequency traded
volume stationary probability distribution of shares in stock markets. The
present scenario, related with superstatistics and thus with non-extensive
statistical mechanics, contains two ingredients that are often considered as
elemental in the dynamics of such complex systems: memory (via multiplicative
noise) and microscopic mechanisms of herding between traders (through the
local temporal fluctuations of $\beta$ or the $v$ mean value) caused, {\it e.g.}, by news, rumours or simply price movements. 
The description of a completely related return-volatility-volume mesoscopic dynamical theory, within
non-extensive statistical mechanics framework, for financial markets is currently 
in progress.

\bigskip

SMDQ would like to thank to C. Tsallis for his continuous incentive and relevant conversations and suggestions; E.P. Borges is 
acknowledged for preliminary discussions about PDF form (\ref{prob-obt}) possible origins and comments made 
on previous versions of this manuscript as well as L.G. Moyano. This work was developed with infrastructural 
support from CNPq and PRONEX/MCT (Brazilian agencies) and financial support from FCT/MCES 
(Portuguese agency; contract SFRH/BD/6127/2001).

%\end{multicols}

\end{document}